\documentclass[a4paper,11pt]{article}
\usepackage{jheppub} % for details on the use of the package, please see the JINST-author-manual
\usepackage{lineno}
\usepackage{subcaption}
\usepackage{graphicx}
\usepackage{tikz-feynman}
\title{\boldmath Dark Matter signals in solar neutrinos fluxes as probe of non-linear symmetry breaking}
\author{A. Carrillo-Monteverde\footnote{Corresponding author.}}
\author{and L. López-Lozano }
\affiliation{\'Area Acad\'emica de Matem\'aticas y F\'isica, Universidad Aut\'onoma del Estado de Hidalgo, Carr. Pachuca-Tulancingo Km. 4.5, C.P. 42184, Pachuca, Hidalgo, M\'exico}
\emailAdd{alba\_carrillo@uaeh.edu.mx}

\abstract{Dark matter (DM) particles gravitationally captured by the Sun can accumulate in its core and subsequently annihilate, producing neutrino fluxes that may be detectable on Earth. The intensity of these fluxes is highly sensitive to the properties of the underlying DM model, especially when the DM candidate is a scalar particle originating from spontaneous or non-linear symmetry breaking mechanisms. In this work, we explore the potential of solar neutrino fluxes to distinguish between the Standard Model extended by a scalar singlet and the non-linear Higgs portal scenarios in the context of a future DM discovery. We compute the expected neutrino fluxes within the regions of parameter space consistent with both relic density and current direct detection limits. Our results show that the non-linear model predicts neutrino fluxes that are systematically larger than those of the linear case—typically by at least one order of magnitude, and up to six orders of magnitude for DM masses around 1 TeV. These findings suggest that solar neutrino observations could provide a valuable probe to discriminate between these competing dark matter frameworks.}

\begin{document}
\maketitle
\flushbottom

\section{Introduction}
\label{sec:intro}
The discovery of the Higgs boson in 2012 \cite{ATLAS:2012yve,CMS:2012qbp,CMS:2013btf} experimentally completed the Standard Model (SM), establishing it as the most successful description of fundamental physics at the electroweak scale. In this framework, the Higgs mechanism employs scalar doublets under the $SU(2)_L$ gauge symmetry to generate mass terms for fermions and electroweak gauge bosons through spontaneous symmetry breaking (SSB) of the $SU(2)_L \times U(1)_Y$ gauge group\cite{Englert1964,Higgs1964,Higgs1964b}.

However, fundamental questions remain regarding the elementary nature of the Higgs boson across all energy scales. Theoretical consistency does not necessarily require the Higgs to be a truly elementary particle, and it may possess internal structure that becomes manifest only at very high energy scales beyond the reach of current LHC explorations. Alternative theoretical frameworks propose that electroweak symmetry breaking occurs dynamically through strong interactions, as exemplified by Composite Higgs Models (CHM) \cite{Panico2015,Chung2021,Chivukula1987,Gripaios2009, Kaplan1984a,Cheng2020}. These models implement dynamical symmetry breaking (DSB) without relying on fundamental scalar fields, instead generating the observed phenomenology through the formation of bound states and condensates arising from strong dynamics \cite{Kaplan1984a,Kaplan1984b}.

The experimental evidence for Dark Matter (DM) at astrophysical scales strongly suggests the existence of New Physics (NP) beyond the SM \cite{ARBEY2021103865}. However, to date, no evidence has been found for direct interactions between DM and SM particles in terrestrial detection experiments.
In this context, the SM can be viewed as an effective theory emerging as the low-energy limit of a more fundamental framework that preserves SM predictions at the electroweak scale. Such extensions might accommodate DSB scenarios compatible with the experimental measurements of Higgs boson interactions observed at the LHC.
Although numerous approaches exist for extending the SM to incorporate DM effects, the Higgs portal \cite{Arcadi:2019lka} has emerged as one of the most extensively studied frameworks in the literature \cite{Craig2016,Djouadi2013,March-Russell2008,Bishara2016,Lebedev2012,Fedderke2014,Lopez-Honorez2012,Andreas2008,Patt2006,Englert2011,Silveira1985,Greljo2013}. At low energies, Higgs portal models with nonlinear symmetry breaking exhibit distinct phenomenology compared to their linear counterparts, providing observables that can differentiate between these scenarios. A key distinction is that nonlinear Higgs portal models typically contain additional parameters, which can help evade the exclusion limits that constrain linear models when relic abundance and direct detection experimental bounds are imposed.

The Nonlinear Higgs Portal (NLHP) \cite{Brivio2016} is a generic framework representing the low-energy limit of models such as CHM, formulated in terms of Higgs Effective Field Theory (HEFT) \cite{Brivio2019} coupled to a singlet scalar DM candidate. The key feature of these effective theories is the potential for mixing between the Higgs boson and new degrees of freedom, enabling small deviations from the standard $SU(2)$ doublet structure. At the effective level, the Higgs boson enters the effective operators as a singlet field.
Within this parametrization, DM coupling to SM particles can produce highly energetic neutrinos detectable by neutrino telescopes. The search for DM annihilation signals in neutrino telescopes represents a well-established approach that has been extensively studied \cite{Silk1985,Press1985,Griest1987,Srednicki1987,Gould1987,Gould1987b,Gould1992,Bernal2013,Catena2015,Bell2021,Chauhan2024}, originating from early attempts to resolve solar neutrino anomalies. Recent advances in solar modeling and sophisticated computational methods for solar neutrino flux calculations (see \cite{Xu2023} and references therein) have enhanced the signal-to-background discrimination capabilities of dedicated neutrino telescopes for detecting beyond-standard-model physics. Combined constraints from relic abundance measurements \cite{planck2023}, gamma-ray observations \cite{GammaRays}, and neutrino flux data \cite{neutrinosun} provide complementary probes of DM interactions through indirect detection channels.
The absence of positive signals in direct detection experiments has imposed stringent constraints on DM mass and coupling parameters, approaching the neutrino floor sensitivity limit \cite{NeutrinoFloor}. This situation underscores the growing importance of refined indirect detection strategies. While current experimental measurements of neutrino fluxes remain insufficient to definitively determine the high-energy symmetry structure of the underlying model, they could potentially provide sufficient discrimination power to distinguish between SSB and DSB mechanisms for EWSB, assuming the existence of a Higgs portal interaction.

In this paper, we employ the NLHP model to parameterize potential deviations from standard electroweak symmetry breaking. Our central hypothesis is that solar neutrino fluxes may contain signatures of non-linear, decorrelated DM particle annihilation processes. We begin by calculating the annihilation rate of a scalar DM candidate into pairs of SM particles. Subsequently, we determine the corresponding neutrino fluxes using simulations from the PPPC4 collaboration \cite{pppc4neutrinos}. We demonstrate that the nonlinear Higgs interactions exhibit distinct behavior compared to the standard SM + singlet model (SM+Scalar) \cite{Cline2013}, which serves as our reference framework, while remaining consistent with updated Planck relic abundance measurements. Additionally, we present current constraints on the parameter space from direct detection experiments for both theoretical frameworks and calculate the branching ratios for all relevant channels within the allowed regions of masses and couplings.
Using this methodology, we find that the allowed parameter region for the SM + singlet model is severely constrained and can be clearly distinguished from the NLHP predictions.
The paper is organized as follows. Section \ref{NLHPModel} explains the nature of the NLHP as an extension of HEFT and presents the primary sectors that generate neutrino fluxes from DM annihilation through Higgs portal processes of the form $SS \to h$ and $SS \to hh$. Section \ref{NeutrinoFlux} details the calculation of solar neutrino fluxes for various DM candidate masses and other free parameters. Finally, Section \ref{conclusions} presents our conclusions.

\section{The scalar sector of the Non Linear Higgs Portal Model versus Linear Higgs Portal}\label{NLHPModel}
In this section, we analyze only those aspects of the NLHP that are relevant for distinguishing between nonlinear and linear Higgs portals. The NLHP encompasses a well-defined subset of operators within the Higgs Effective Field Theory (HEFT) framework that parameterizes New Physics contributions to the scalar sector. Following the approach of $\chi$PT, HEFT introduces three Goldstone bosons via the standard operator $U=\exp{(i \tau^a \pi^a /v)}$, which transforms under $SU(2)_L\times SU(2)_R$ as $U'=LUR^\dagger$, where $v$ is a characteristic scale that suppresses higher-order contributions. In the context of Composite Higgs Models (CHM), the $\pi^a$ represent degrees of freedom associated with a heavier mass scale, making $v$ a scale that exceeds the cutoff of the effective theory.

The NLHP parameterize the low-energy regime of different models, for instance composite Higgs models, where scalar interactions are not described by conventional doublets just above the electroweak scale, but rather through non-renormalizable power functions of the Higgs boson $h$. The interactions between Standard Model degrees of freedom and the Higgs boson emerge as a consequence of the $SU(2)_L \times SU(2)_R$ breaking, which approximately encompasses all Standard Model couplings. This symmetry breaking proceeds hierarchically as:
\begin{equation}
SU(2)_L\times SU(2)_R \to SU(2)_C\to U(1)_Y.
\end{equation}

At the scale where left-right symmetry becomes manifest, scalars enter the model as pseudo Nambu-Goldstone (pNG) bosons, which generate nonlinear interactions with fermions at low energies. Specifically, the Higgs boson is incorporated into the Lagrangian through functional forms that enable a consistent perturbative treatment \cite{Brivio2016} within the HEFT framework \cite{Brivio2019}.

The only experimentally observed scalar particle is the Higgs boson with a mass of $m_h \simeq 125\,\text{GeV}$ at the electroweak scale. Within this framework, the non-active pNG bosons are integrated out, with the exception of one dark matter candidate (S). The scalar sector of the NLHP is consequently given by:

\begin{equation}
    \mathcal{L}_S=\frac{1}{2}\partial_\mu S\partial^\mu S-\frac{m^2_S}{2}S^2\mathcal{F}_{S_1}(h)-\lambda S^4\mathcal{F}_{S_2}(h)+\sum_{i=1}^5c_i\mathcal{A}_i(h).
\end{equation}
where the high-dimensional effective operators involving only scalars have been introduced with the following operators:
\begin{eqnarray}
\mathcal{A}_1(h)&=&\text{Tr}(\mathbf{V}_\mu\mathbf{V}^\mu)S^2\mathcal{F}_1(h),\\
\mathcal{A}_2(h)&=&S^2\Box\mathcal{F}_2(h),\\
\mathcal{A}_3(h)&=&\text{Tr}(\mathbf{T}\mathbf{V}_\mu)\text{Tr}(\mathbf{T}\mathbf{V}^\mu)S^2\mathcal{F}_3(h),\\
\mathcal{A}_4(h)&=&i\text{Tr}(\mathbf{T}\mathbf{V}_\mu)(\partial^\mu S^2)\mathcal{F}_4(h),\\
\mathcal{A}_5(h)&=&i\text{Tr}(\mathbf{T}\mathbf{V}_\mu) S^2\partial^\mu\mathcal{F}_5(h).
\end{eqnarray}
The contribution of the pseudo Nambu-Goldstone Bosons (pNGB) has been introduced in the usual way, involving scalars and vector fields of the form:
\begin{equation}
\mathbf{T}=\mathbf{U}\sigma^3\mathbf{U}^\dagger\quad;\quad \mathbf{V}_\mu=(\mathbf{D}_\mu\mathbf{U})\mathbf{U}^\dagger,
\end{equation}
where $\mathbf U=\exp{(i\sigma_a\pi^a/v)}$ and the covariant derivative
\begin{equation}
\mathbf D_\mu=\partial_\mu \mathbf U+ig W^a_\mu\sigma_a \mathbf U-\frac{ig'}{2}B_\mu\mathbf U \sigma_3.
\end{equation}
The dependence on the Higgs boson is introduced through power functions of $h$ as
\begin{eqnarray}
	\mathcal{F}_{S_j}(h)&=&1+2a_{S_j}\frac{h}{v}+b_{S_j}\frac{h^2}{v^2}+\mathcal{O}\left(\frac{h^3}{v^3}\right)\label{Fsfunction},\\
	\mathcal{F}_i(h)&=&1+2a_i\frac{h}{v}+b_i\frac{h^2}{v^2}+\mathcal{O}\left(\frac{h^3}{v^3}\right)\label{Fsfunction2}.
\end{eqnarray}
The series is truncated at the terms that generate the correct dimension according to the chosen scale. The introduction of (\ref{Fsfunction}) allows for the removal of the correlation between the $SS \to h$ and $SS \to hh$ decays by introducing new parameters, $c_1$ and $c_2$, in a minimal scenario. The interaction of the Higgs boson with vector bosons arises from the electroweak sector via the covariant derivative, with the terms of the Lagrangian given by
\begin{equation}
\mathcal{L}_\text{EW}\supset -\frac{v^2}{4}\text{Tr}(\mathbf V_\mu\mathbf V^\mu)\mathcal{F}_C(h)+c_T\frac{v^2}{4}\text{Tr}(\mathbf T \mathbf V_\mu)\text{Tr}(\mathbf T \mathbf V^\mu)\mathcal F_T(h).
\end{equation}
The functions $\mathcal{F}_j(h)$ for $j = C, T$ have a form similar to (\ref{Fsfunction2}), involving powers of $h$ to obtain SM interactions between the Higgs boson and vector bosons. Here, we are interested only in the interactions that generate neutrinos in the final states to calculate the fluxes coming from the Sun.

\section{The signals of Dark Matter in neutrino telescopes}\label{NeutrinoFlux}

The Sun serves as a well-understood and precisely modeled source of neutrinos. Contemporary Standard Solar Models (SSMs) yield accurate and reliable predictions for the energy spectra and fluxes of neutrinos arising from thermonuclear processes in the solar core. This high level of theoretical precision enables the robust identification of possible non-standard contributions to the solar neutrino flux, particularly in scenarios where neutrino masses and couplings permit energy spectra that deviate from or exceed those predicted by the SSM.

In particular, neutrino signals arising from dark matter (DM) capture and annihilation in the Sun have been extensively studied as a promising indirect detection channel \cite{captureSunWIMPS}. Although the overall contribution of dark matter annihilation to the total neutrino flux is expected to be small for DM conventional models, the energy spectra of these neutrinos can differ significantly from those predicted by the SSM, particularly when dark matter is modeled as weakly interacting massive particles (WIMPs) \cite{WimpDM}. 

Non-standard neutrino fluxes from the Sun have also been investigated in a variety of theoretical frameworks, including models with Higgs portal interactions \cite{nuHiggsPortal}. Building on this context, we explore the potential of solar neutrino fluxes as probes of dark matter interactions in two specific scenarios: the Standard Model plus Scalar Dark Matter (SM+Scalar) and the Non-Linear Higgs Portal (NLHP). Our objective is to determine whether the neutrino signals associated with these models can provide indirect evidence of dark matter and whether they can help discriminate between minimal and non-linear symmetry-breaking scenarios.

The first step in our study involved updating the viable parameter space for both models using the most recent constraints from cosmological and direct detection data. We implemented the SM+Scalar and NLHP models in FeynRules \cite{feynrules} to generate the necessary model files and performed detailed calculations using MicrOmegas \cite{micromegas}. We computed the relic density and direct detection cross-sections, applying the latest exclusion limits from Planck \cite{planck2023} and Xenon1T \cite{XENON:2018voc} to identify the allowed regions in the $m_{DM}$ versus $\lambda_{SH}$ parameter space, considering in the case of the NLHP, the contribution of each additional effective term in the potential singly.

\begin{figure}[htbp]
\centering
\includegraphics[width=.9\textwidth]{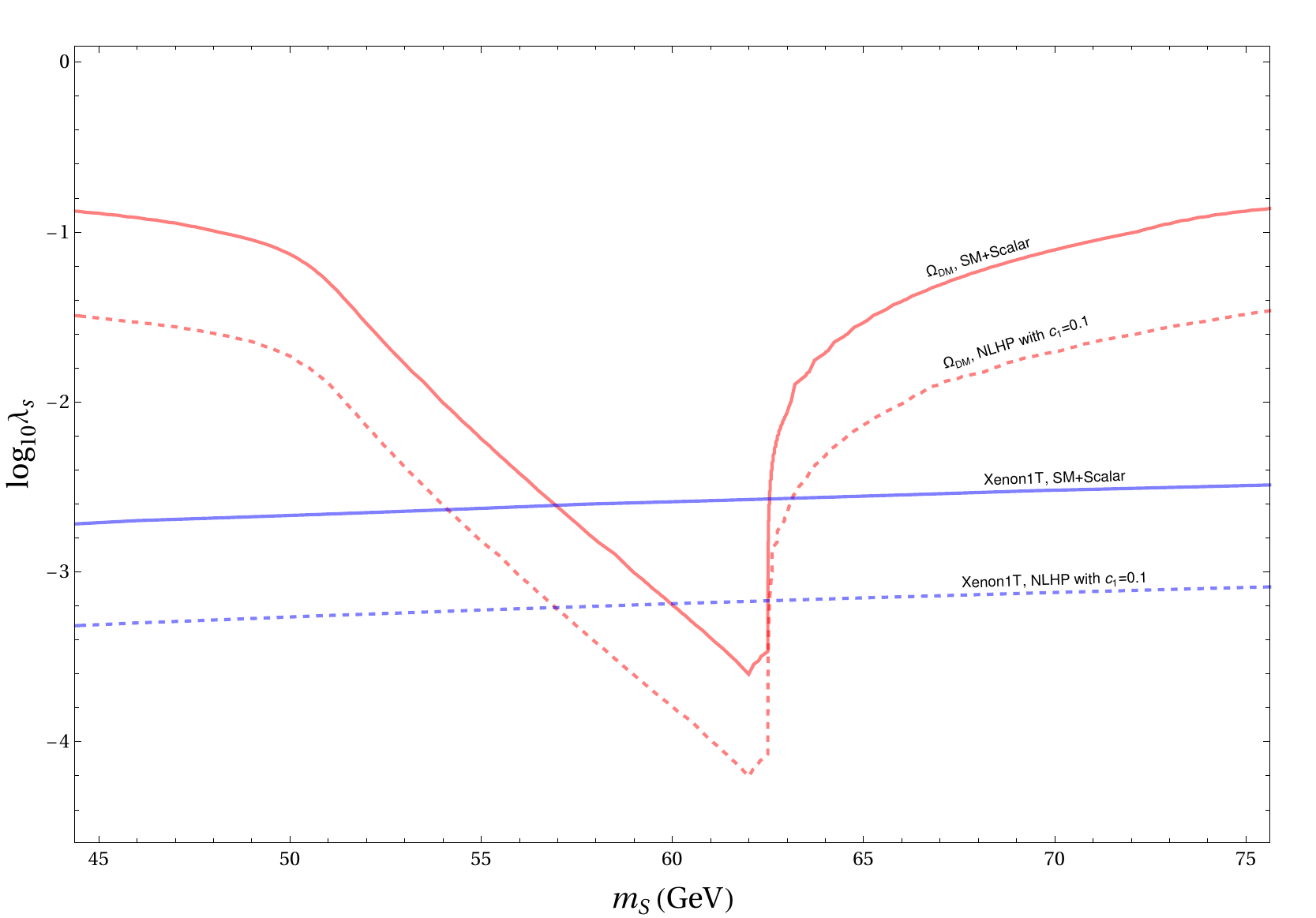}
\caption{The allowed region for the parameters in the SM+Scalar and NLHP with $c_1=0.1$ is shown. The restrictions arise from direct detection limits \cite{XENON:2018voc} and relic abundance \cite{planck2023}. The allowed region in each case is the area corresponding to the (dotted or solid) blue line from below and the (dotted or solid) red line from above.\label{fig:1a}}
\end{figure}

\begin{figure}[htbp]
\centering
\includegraphics[width=.9\textwidth]{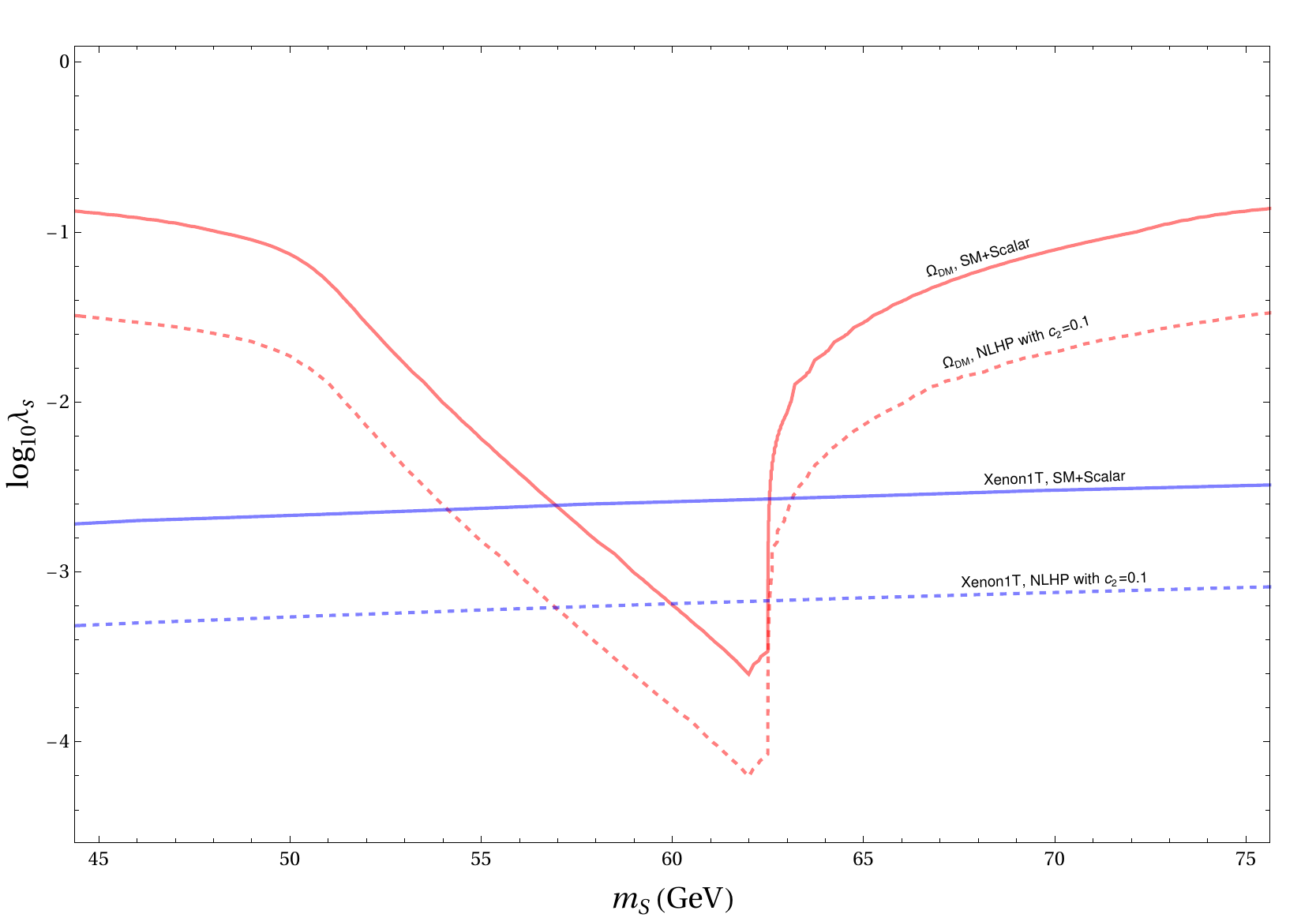}
\caption{Allowed region for the parameters in the SM+Scalar and NLHP with $c_2=0.1$ is shown. Restrictions arise from direct detection limits \cite{XENON:2018voc} and relic abundance \cite{planck2023}. The allowed region in each case is the area corresponding to the (dotted or solid) blue line from below and the (dotted or solid) red line from above. \label{fig:1b}}
\end{figure}

Figures \ref{fig:1a} and \ref{fig:1b} show the updated parameter spaces for both models under consideration. These results confirm that although most of the low-mass region has been excluded by current experimental constraints, a small viable region remains for dark matter masses below 100 GeV in both frameworks. This residual window motivated a detailed analysis, with particular emphasis on the case of $m_{DM} \simeq 60$ GeV, which continues to be a phenomenologically relevant mass scale compatible with present bounds.

In these figures, solid lines represent the Standard Model plus scalar singlet (SM+Scalar) scenario, while dashed lines correspond to the non-linear Higgs portal (NLHP) model. For the relic density constraint, the allowed parameter space lies above the corresponding curve, whereas for direct detection, the allowed region lies below the respective exclusion limit. Consequently, the viable region for each model is defined by the intersection of these conditions: it must lie above the direct detection limits and below the relic density curve.

Notably, both models remain compatible within the same parameter space region, as delineated by the experimental constraints in the SM+Scalar scenario. This overlapping region serves as the basis for further comparison of the predicted neutrino fluxes.

\begin{figure}\centering

\begin{subfigure}{0.4\textwidth}
\begin{tikzpicture}
\begin{feynman}
\coordinate (i1) at (0,2);
\coordinate (i2) at (0,0);
\coordinate (a) at (1,1);
\coordinate (b) at (3,1);
\coordinate (f1) at (4,2);
\coordinate (f2) at (4,0);
\node [above right] at (i1) {\(S\)};
\node [below right] at (i2) {\(S\)};
\node [below left] at (f2) {\(\nu,\overline{\nu}\)};
\draw[thick] (3.1, 0.9) -- ( 4.1, 1.9);
\draw[thick] (2.9, 1.1) -- ( 3.9, 2.1);
\node[above right] at (f1) {\(X\)};
\diagram*[layered layout]{
{(i1), (i2)} -- [scalar,very thick] (a) -- [fermion,half left,very thick,edge label=\(f\)] (b) -- [fermion,half left,very thick,edge label=\(\overline{f}\)] (a),
(b) -- [fermion, thick] (f2),
(b) -- [thick] (f1),
};
\filldraw [fill=gray!40, very thick] (a) circle (9pt) node[font=\small] {NP};
\filldraw [fill=gray!40, very thick] (b) circle (9pt) node[font=\small] {SM};
\end{feynman}
\end{tikzpicture}
\caption{}
\label{fig0}
\end{subfigure}
%%%%%%%%%%%%%%%%%%%%%%%%%%%%%%%%%%%%%
\begin{subfigure}{0.4\textwidth}
\begin{tikzpicture}
\begin{feynman}
\coordinate (i1) at (0,2);
\coordinate (i2) at (0,0);
\coordinate (a) at (1,1);
\coordinate (b) at (3,1);
\coordinate (f1) at (4,2);
\coordinate (f2) at (4,0);
\node [above right] at (i1) {\( S\)};
\node [below right] at (i2) {\(S\)};
\node [below left] at (f2) {\(\nu,\overline{\nu}\)};
\draw[thick] (3.1, 0.9) -- ( 4.1, 1.9);
\draw[thick] (2.9, 1.1) -- ( 3.9, 2.1);
\node[above right] at (f1) {\(X\)};
\diagram*[layered layout]{
{(i1), (i2)} -- [scalar,very thick] (a) -- [boson,very thick,half left,edge label={\(W,Z^{0}\)}] (b) -- [boson,very thick,half left,edge label={\(W,Z^{0}\)}] (a),
(b) -- [thick] (f2),
(b) -- [thick] (f1),
};
\filldraw [fill=gray!40, very thick] (a) circle (9pt) node[font=\small] {NP};
\filldraw [fill=gray!40, very thick] (b) circle (9pt) node[font=\small] {SM};
\end{feynman}
\end{tikzpicture}
    \caption{}
    \label{fig1}
\end{subfigure}
%%%%%%%%%%%%%%%%%%%%%%%%%%%%%%%%%%%%%%%%%%
\begin{subfigure}{0.4\textwidth}
\begin{tikzpicture}
\begin{feynman}
\coordinate (i1) at (0,2);
\coordinate (i2) at (0,0);
\coordinate (a) at (1,1);
\coordinate (b) at (3,1);
\coordinate (f1) at (4,2);
\coordinate (f2) at (4,0);
\node [above right] at (i1) {\(S\)};
\node [below right] at (i2) {\(S\)};
\node [below left] at (f2) {\(\nu,\overline{\nu}\)};
\draw[thick] (3.1, 0.9) -- ( 4.1, 1.9);
\draw[thick] (2.9, 1.1) -- ( 3.9, 2.1);
\node[above right] at (f1) {\(X\)};
\diagram*[layered layout]{
{(i1), (i2)} -- [scalar,very thick] (a) -- [scalar,half left,very thick,edge label=\(h^{0}\)] (b) -- [scalar,half left,very thick,edge label=\(h^{0}\)] (a),
(b) -- [fermion,thick] (f2),
(b) -- [thick] (f1),
};
\filldraw [fill=gray!40, very thick] (a) circle (9pt) node[font=\small] {NP};
\filldraw [fill=gray!40, very thick] (b) circle (9pt) node[font=\small] {SM};
\end{feynman}
\end{tikzpicture}
\caption{}
\label{fig2}
\end{subfigure}
%%%%%%%%%%%%%%%%%%%%%%%%%%%%%%%%%%%%%%%%%%
\begin{subfigure}{0.4\textwidth}
\begin{tikzpicture}
\begin{feynman}
\coordinate (i1) at (0,2);
\coordinate (i2) at (0,0);
\coordinate (a) at (1,1);
\coordinate (b) at (3,1);
\coordinate (f1) at (4,2);
\coordinate (f2) at (4,0);
\node [above right] at (i1) {\(S\)};
\node [below right] at (i2) {\(S\)};
\node [below left] at (f2) {\(\nu,\overline{\nu}\)};
\draw[thick] (3.1, 0.9) -- ( 4.1, 1.9);
\draw[thick] (2.9, 1.1) -- ( 3.9, 2.1);
\node[above right] at (f1) {\(X\)};
\diagram*[layered layout]{
{(i1), (i2)} -- [scalar,very thick] (a) -- [boson,very thick,half left,edge label={\(Z^{0}\)}] (b) -- [scalar,half left,very thick,edge label=\(h^{0}\)] (a),
(b) -- [fermion,thick] (f2),
(b) -- [thick] (f1),
};
\filldraw [fill=gray!40, very thick] (a) circle (9pt) node[font=\small] {NP};
\filldraw [fill=gray!40, very thick] (b) circle (9pt) node[font=\small] {SM};
\end{feynman}
\end{tikzpicture}
\caption{}
\label{fig3}
\end{subfigure}
\label{diagramas}
\caption{Here, the two steps to calculate the neutrino flux production from DM annihilation for each channel of the portal are shown schematically. The NP vertex represents the contribution of the Higgs portal to the generation of SM particles and is calculated using the MicrOmegas software. The SM vertex represents all processes that generate neutrinos at secondary vertices, and these were simulated using data from the PPPC4 collaboration. For details about the SM vertex, see the discussion in the text.}
\end{figure}
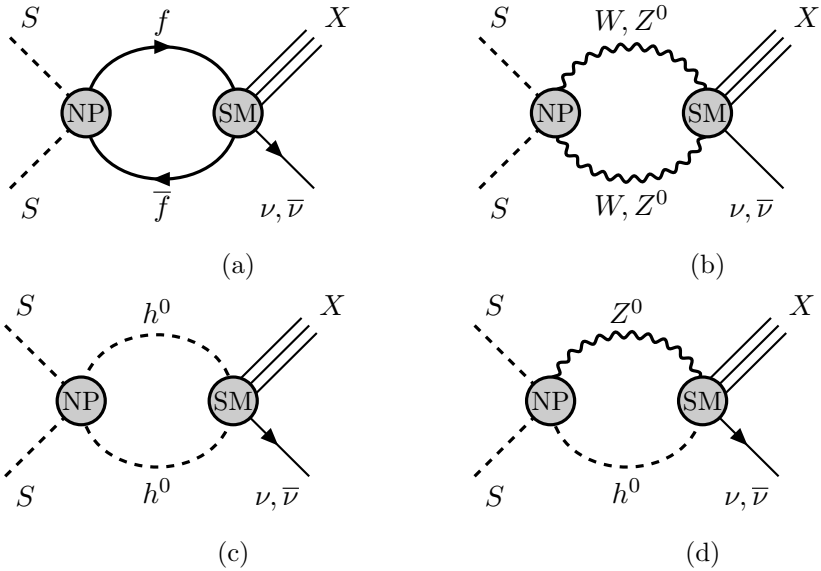

The neutrino fluxes resulting from dark matter annihilation in the Sun were calculated using the tabulated results provided by the PPPC4 collaboration \cite{pppc4neutrinos}. These resources include the differential neutrino spectra $\frac{dN}{dE}$ at Earth, which account for the effects of neutrino interactions and propagation through the Sun, as well as oscillations during transit to Earth.
The fluxes are provided for a range of dark matter masses and specific Standard Model final states. Figures \ref{fig0}–\ref{fig3} illustrate the Feynman-like diagrams considered in the calculation of dark matter (DM) annihilation into pairs of Standard Model (SM) particles. These SM particles can subsequently interact with the surrounding medium and among themselves, producing secondary low-energy neutrinos through processes involving both hadronic and leptonic channels.

For each allowed point in the parameter space of our models, we computed the dark matter annihilation branching ratios into all relevant final states, including $W^+W^-$, $b\bar{b}$, $\tau^+\tau^-$, $ZZ$, and $t\bar{t}$. These branching ratios were obtained using \texttt{MicrOmegas} \cite{micromegas}. The total neutrino flux at Earth was then calculated by combining the contributions from each annihilation channel according to:
\begin{equation}
\frac{d\Phi_\nu}{dE} = \sum_f BR_f \cdot \left( \frac{\Gamma_A}{4\pi D^2} \cdot \frac{dN_\nu^f}{dE} \right),
\label{eq:neutrinoflux}
\end{equation}
where $BR_f$ is the branching ratio into the final state $f$, $\frac{dN_\nu^f}{dE}$ is the differential neutrino spectrum for channel $f$ obtained from the PPPC4 tables, $\Gamma_A$ is the annihilation rate of dark matter particles in the Sun, and $D$ is the distance from the Sun to the Earth.

The annihilation rate $\Gamma_A$ depends on the balance between the dark matter capture and annihilation processes in the Sun. Assuming capture-annihilation equilibrium, which is generally valid for WIMP candidates over cosmological timescales, the annihilation rate is given by:
\begin{equation}
\Gamma_A = \frac{1}{2} C_\odot,
\label{eq:annihilationrate}
\end{equation}
where $C_\odot$ is the dark matter capture rate in the Sun. The capture rate was calculated using the scattering cross sections provided by the model, the local dark matter density, and the standard Maxwellian velocity distribution. The capture rate computations were performed using \texttt{MicrOmegas}, which includes the necessary nuclear form factors and astrophysical parameters.

Three benchmark dark matter masses were selected: 60 GeV, 500 GeV, and 1 TeV. The 60 GeV case is particularly relevant because of the surviving allowed region, while the higher masses were chosen to explore the model behavior in regions less constrained by direct detection experiments but potentially accessible to neutrino telescopes.

\begin{figure}[htbp]
\centering
\includegraphics[width=.32\textwidth]{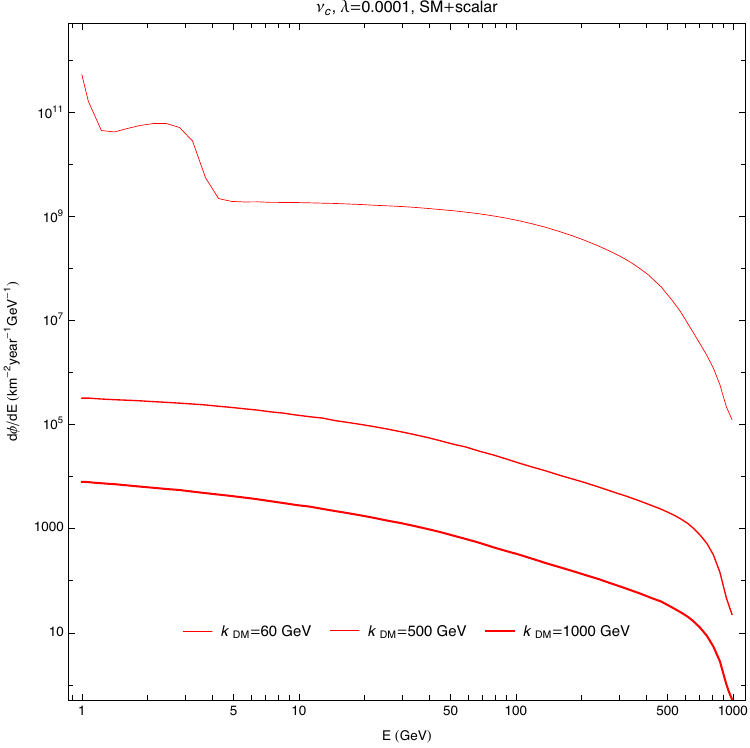}
\includegraphics[width=.32\textwidth]{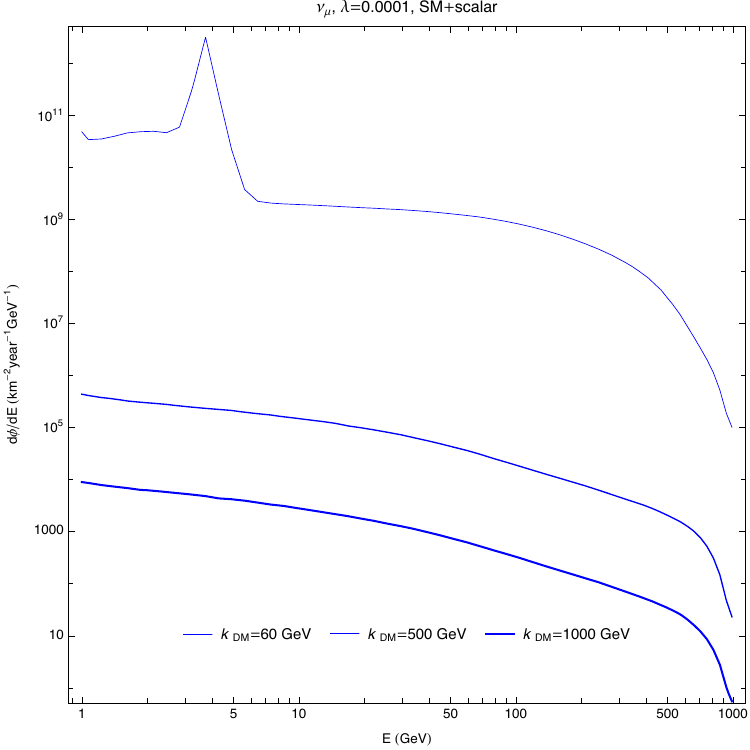}
\includegraphics[width=.32\textwidth]{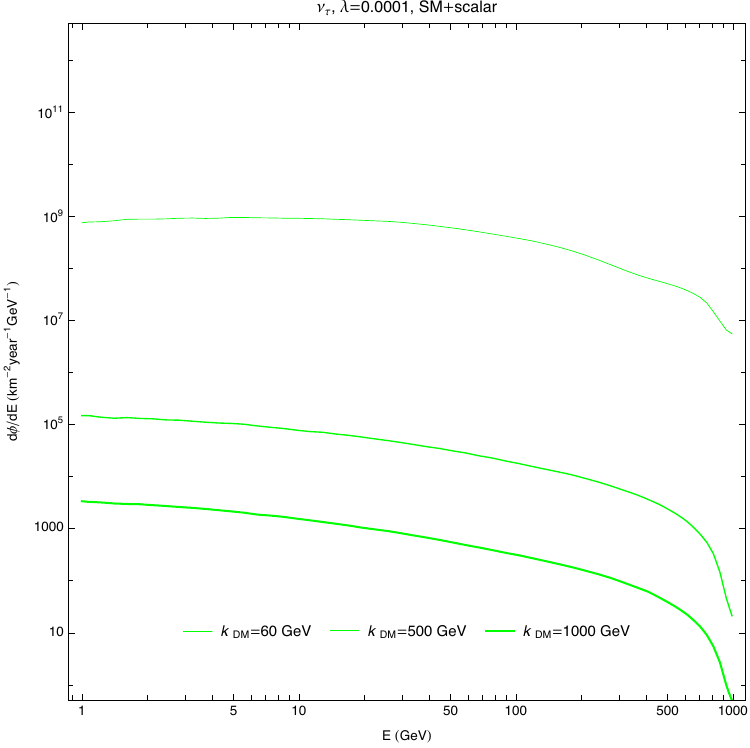}
\caption{Neutrino fluxes using the SM+Scalar model for the three neutrino flavors.\label{fig:3}}
\end{figure}

\begin{figure}[htbp]
\centering
\includegraphics[width=.32\textwidth]{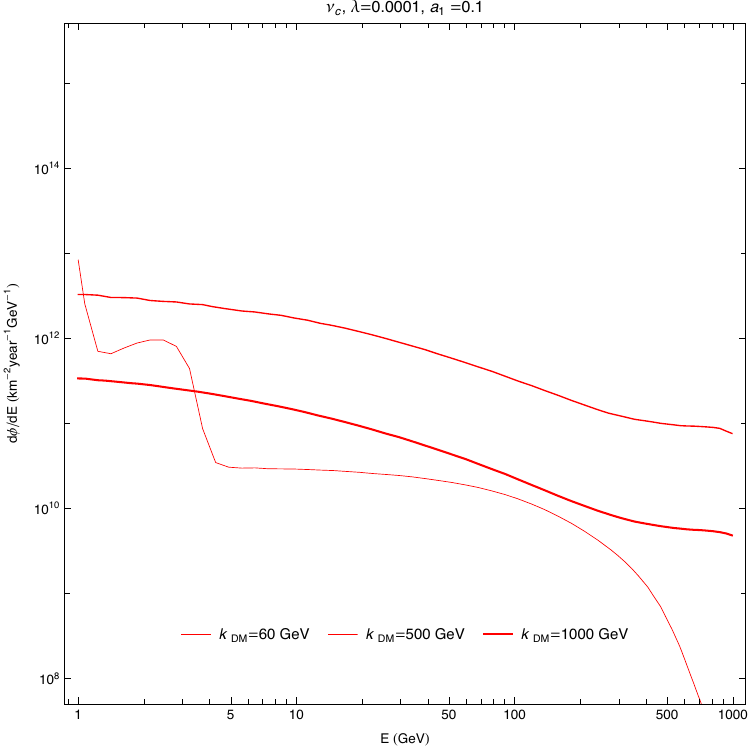}
\includegraphics[width=.32\textwidth]{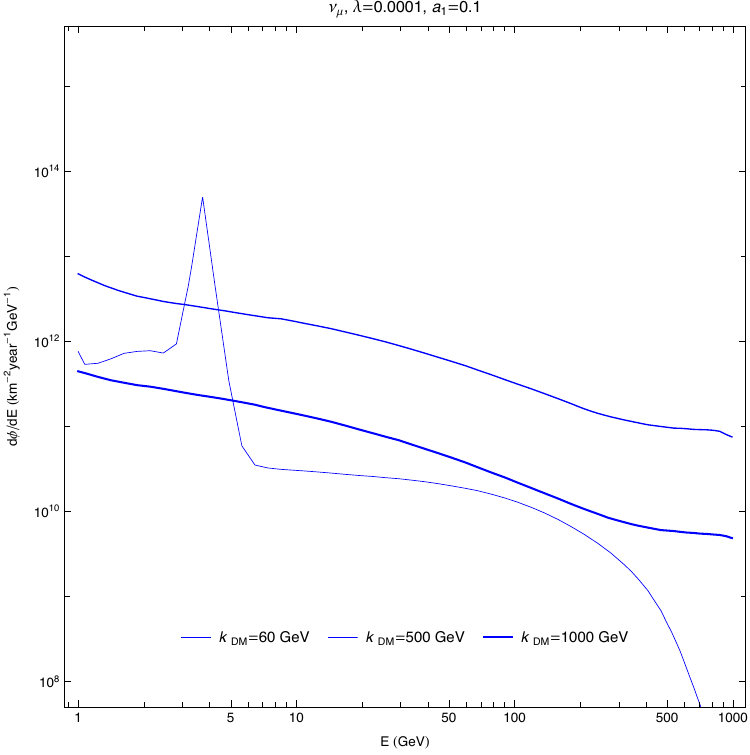}
\includegraphics[width=.32\textwidth]{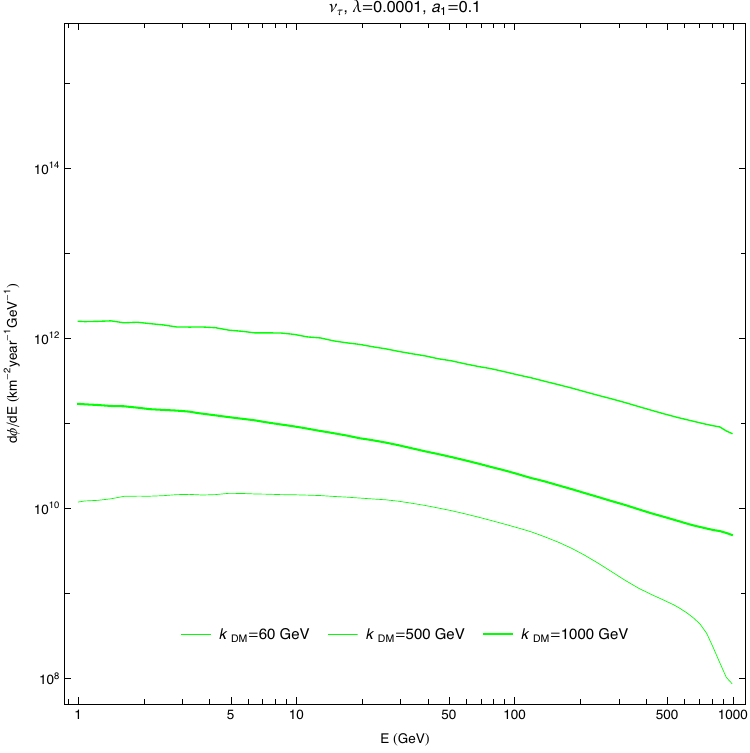}
\\
\includegraphics[width=.32\textwidth]{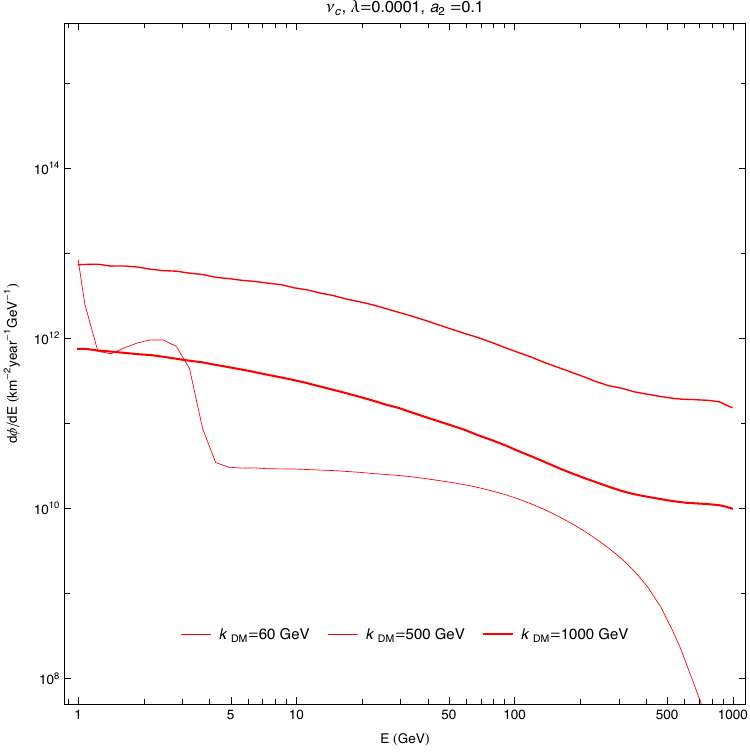}
\includegraphics[width=.32\textwidth]{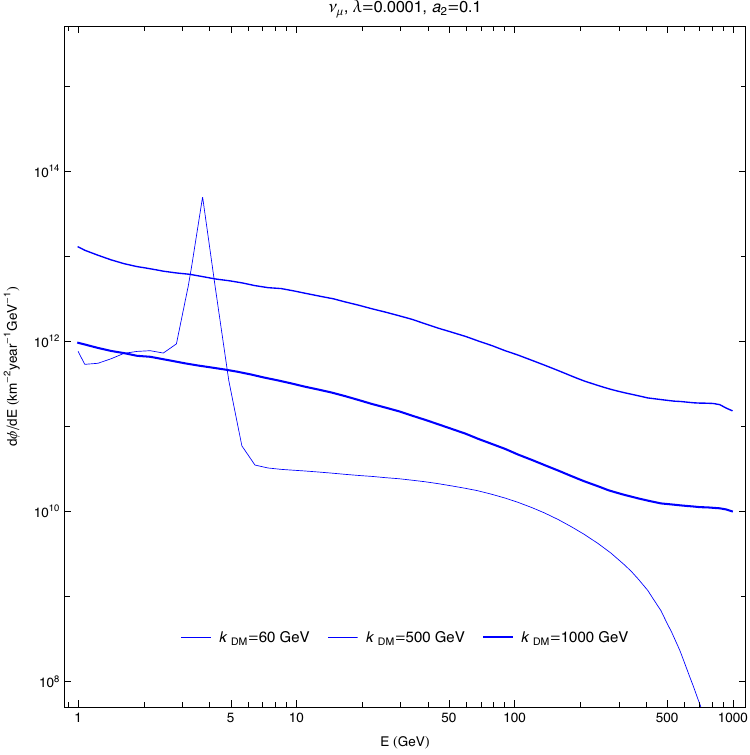}
\includegraphics[width=.32\textwidth]{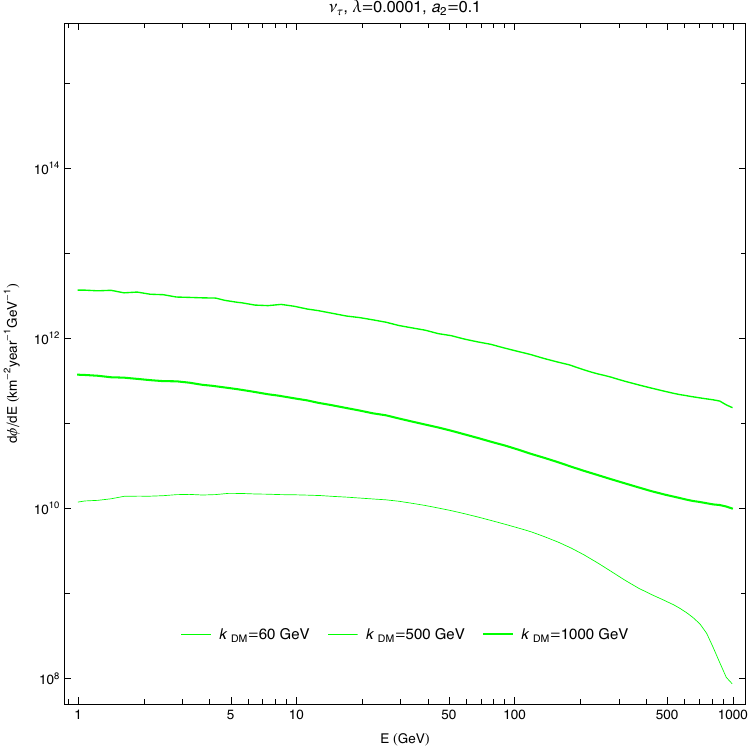}
\caption{Neutrino fluxes for the NLHP model for the three neutrino flavors. On the top panel it's considered a value for the parameter $c_1=0.1$ and the bottom panel a value for $c_2=0.1$ is considered.\label{fig:4}}
\end{figure}

The final neutrino fluxes presented in this work, in Fig \ref{fig:3},  and Fig. \ref{fig:4}, are the sum of the contributions from all annihilation channels weighted by their respective branching ratios and include the full propagation and oscillation effects provided by the PPPC4 framework for the three neutrino flavors. These fluxes are directly comparable to the sensitivity limits of current and future neutrino telescopes such as IceCube and KM3NeT.

\begin{figure}[htbp]
\centering
\includegraphics[width=.95\textwidth]{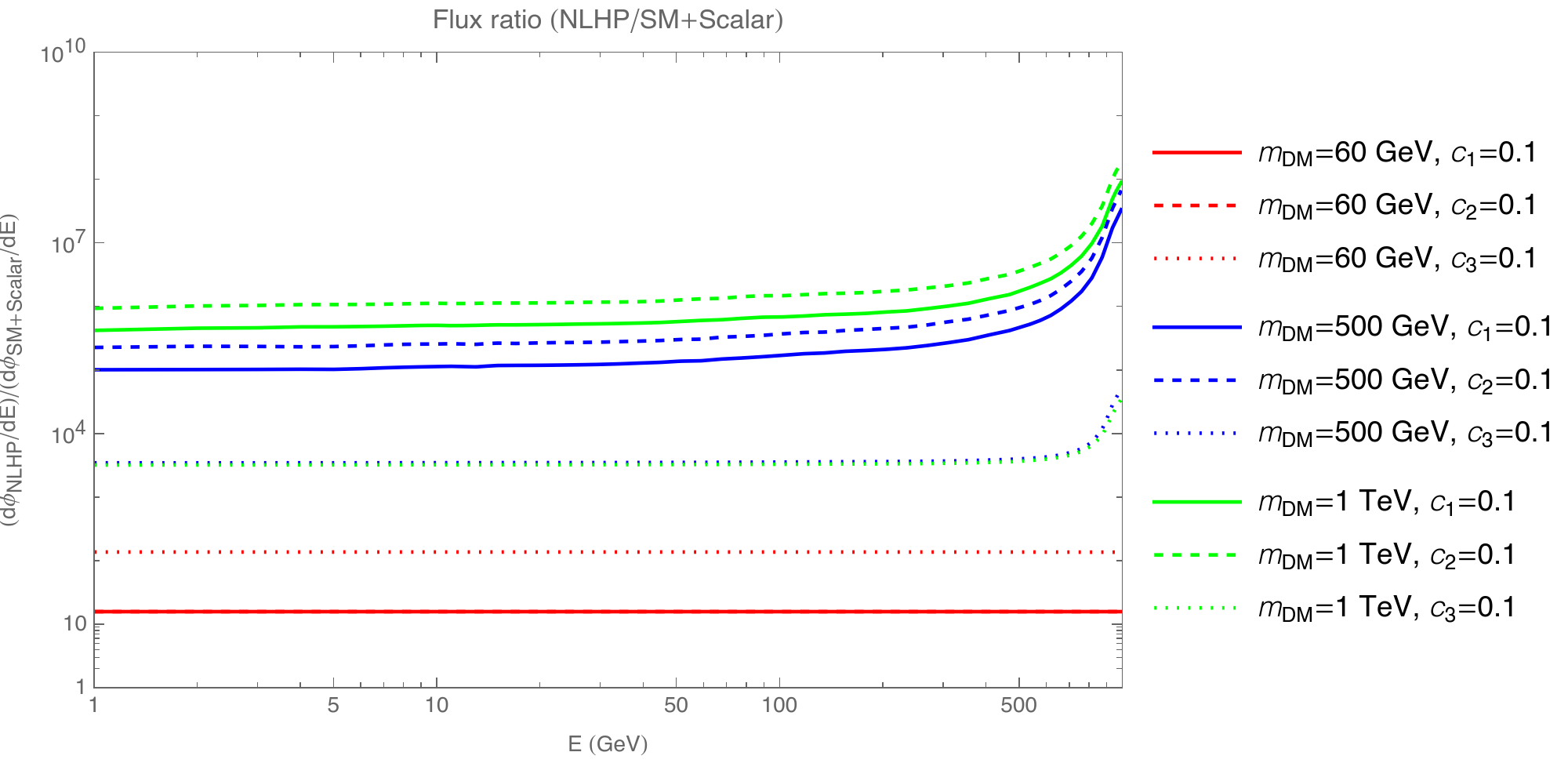}
\caption{Flux ratio for the NLHP model and SM+S.
\label{fig:5}}
\end{figure}

\begin{figure}[htbp]
\centering
\includegraphics[width=.95\textwidth]{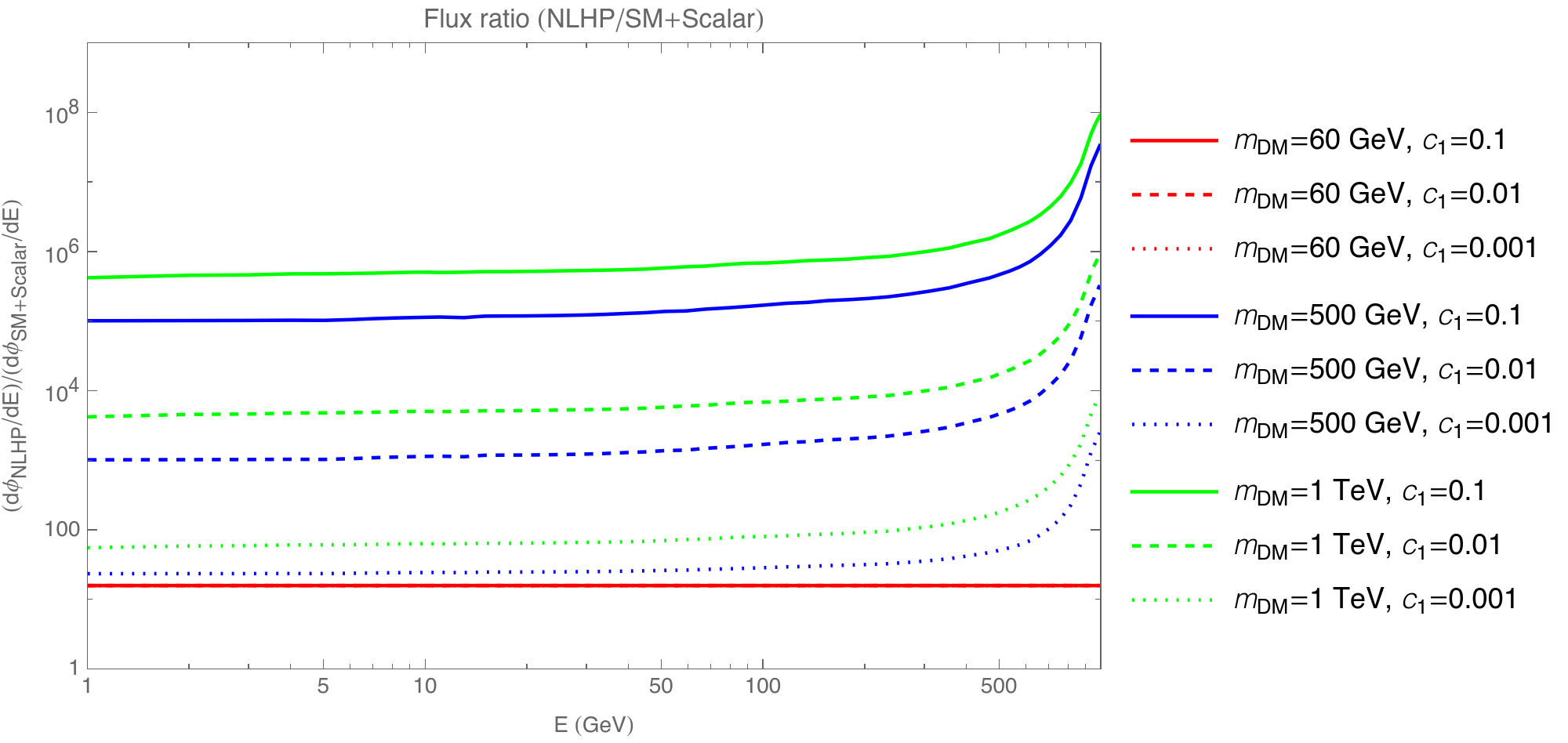}
\caption{Flux ratio for the NLHP model and SM+S with different values of $c_1$.
\label{fig:6}}
\end{figure}

\begin{figure}[htbp]
\centering
\includegraphics[width=.95\textwidth]{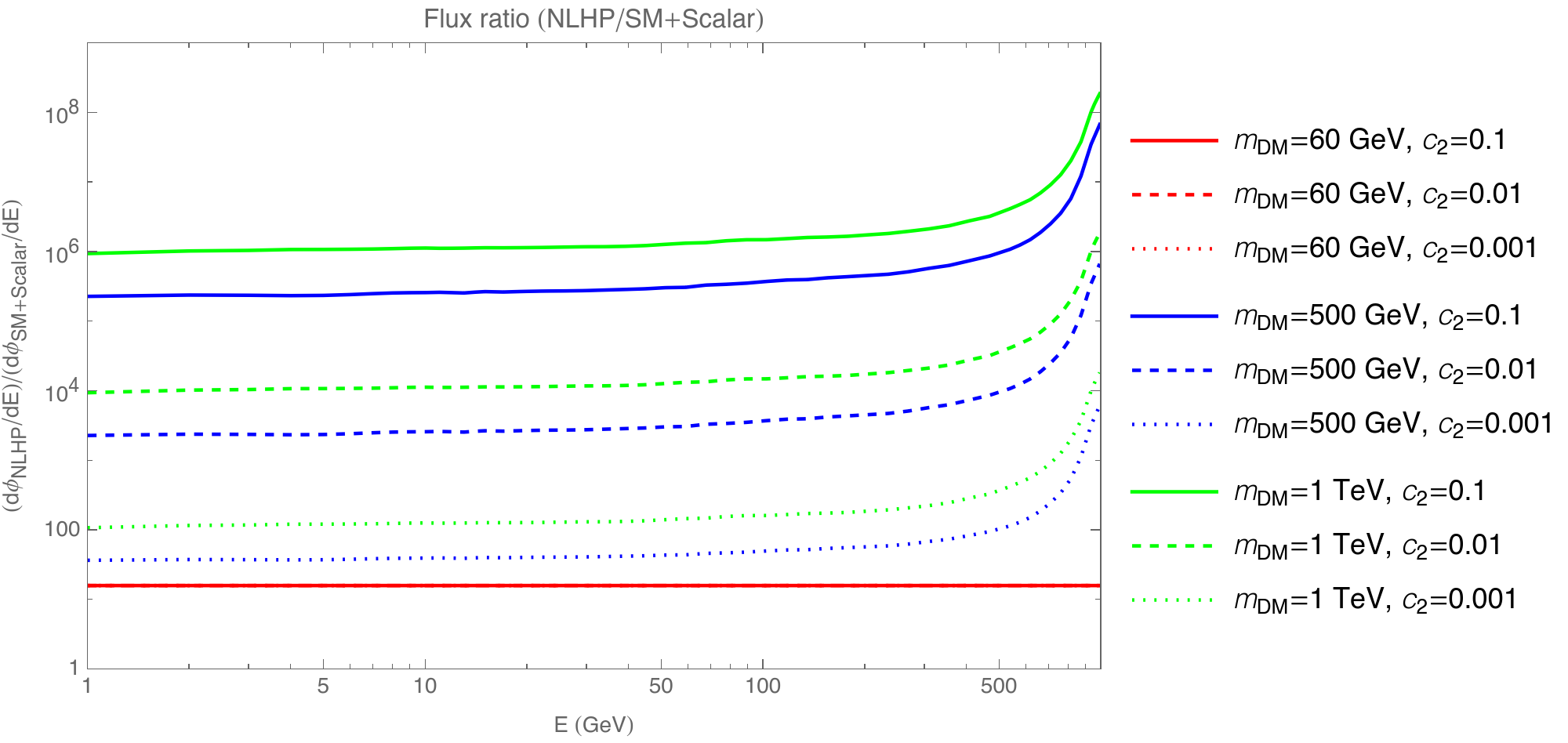}
\caption{Flux ratio for the NLHP model and SM+S with different values of $c_2$.
\label{fig:7}}
\end{figure}

\begin{figure}[htbp]
\centering
\includegraphics[width=.95\textwidth]{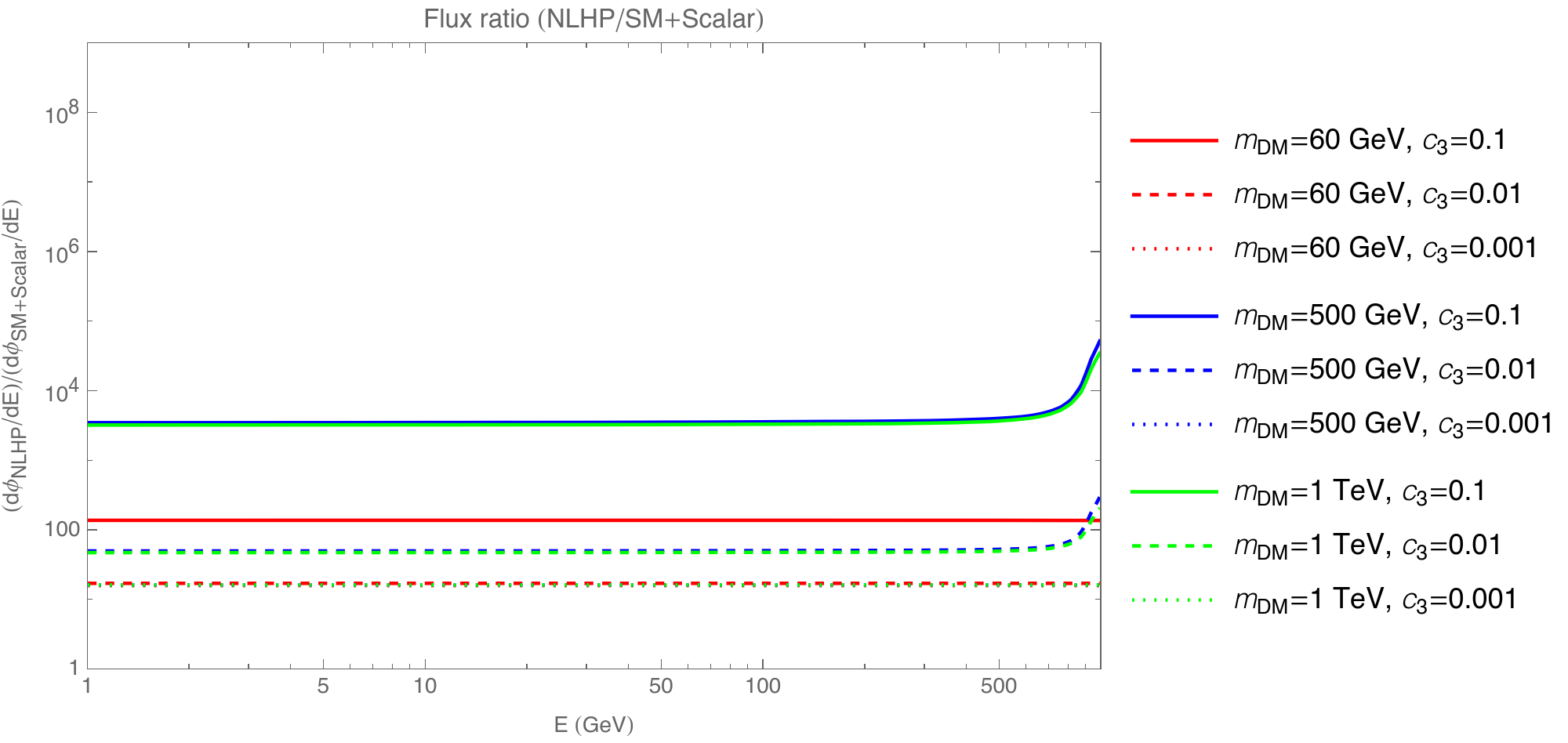}
\caption{Flux ratio for the NLHP model and SM+S with different values of $c_3$.
\label{fig:8}}
\end{figure}

For $m_{DM} = 60$ GeV, we found that the predicted neutrino fluxes from both models are remarkably similar, differing by no more than one order of magnitude, even when varying the additional effective couplings characteristic of the NLHP scenario. In contrast, for $m_{DM} = 500$ GeV and 1 TeV, the flux predictions exhibit significant variation. In these cases, the neutrino fluxes in the NLHP can differ by up to five orders of magnitude depending on the values of the non-linear couplings, as shown in Figures \ref{fig:5}, \ref{fig:6}, \ref{fig:7} and \ref{fig:8} where the ratios of the SM+Scalar neutrino fluxes and NLHP were computed for each value of the DM massed considered and also exploring different order of magnitude in the couplings $c_1$, $c_2$ and $c_3$.

To better understand the impact of the non-linear structure of the NLHP, we performed a detailed parameter scan varying the three additional effective couplings present in this model. Our results show that the non-linear couplings can either enhance or suppress the neutrino fluxes significantly, particularly at high DM masses, where their effects become dominant.

%\begin{figure}[htbp]
%\centering
%\includegraphics[width=.4\textwidth]{plots/%RegionPermitidaNLHp1.pdf}
%\qquad
%\includegraphics[width=.4\textwidth]{plots/%RegionPermitidaNLHp1.pdf}
%\caption{Allowed region for the parameters in the NLHP model is shown in white. On the right is for a parameter $c_1=0.1$ and the plot on the left is with $c_2=0.1$. Restrictions come from direct detection limits and relic abundance as well.\label{fig:2}}
%\end{figure}

The results obtained highlight both the common features and key differences between the SM+Scalar and NLHP models regarding their indirect detection signatures in solar neutrino fluxes.

At low dark matter masses (around 60 GeV), the predicted neutrino fluxes are comparable between the two models. This suggests that, within this mass range, indirect detection via solar neutrinos may not be sufficient on its own to distinguish between the minimal scalar extension and the non-linear portal. However, combining solar neutrino measurements with direct detection constraints and collider searches could still provide valuable complementary information.

In contrast, at higher dark matter masses (500 GeV and 1 TeV), the NLHP model displays a much richer phenomenology. The presence of additional non-linear couplings leads to dramatic variations in the neutrino fluxes, spanning several orders of magnitude. This sensitivity offers a potential avenue to test the non-linear symmetry-breaking structure in future neutrino telescopes. Experiments such as IceCube-Gen2 and KM3NeT may have the capability to probe these enhanced fluxes and provide indirect evidence supporting or constraining NLHP models.

The ability of the NLHP to produce both enhanced and suppressed neutrino fluxes, depending on the choice of effective couplings, underlines the importance of considering non-linear realizations of electroweak symmetry breaking in dark matter phenomenology. These effects would remain invisible in minimal scalar extensions of the SM but could leave distinctive imprints in the high-energy solar neutrino spectrum.

Our results emphasize the complementarity between cosmological, direct, and indirect detection strategies. In particular, the distinct high-mass behavior predicted by the NLHP model could offer a pathway to discriminate between DM scenarios and probe the nature of the underlying symmetry-breaking mechanism.

Future work should aim to include more detailed simulations of neutrino propagation, oscillations, and detector responses to fully assess the observability of these fluxes. Additionally, exploring alternative solar capture rates, DM velocity distributions, and astrophysical uncertainties could further refine the predictions presented here.

\section{Conclusions}\label{conclusions}
In this work, we investigated the potential of solar neutrino fluxes to discriminate between two competing dark matter scenarios: the Standard Model extended by a scalar singlet and the Non-Linear Higgs Portal model. Our analysis focused on the indirect detection of DM through neutrino signals arising from annihilation processes in the Sun, leveraging updated constraints from relic abundance measurements and direct detection experiments.
\begin{itemize}
    \item  The allowed regions for both models are severely restricted by current experimental data, particularly for DM masses below $100\,\mathrm{GeV}$. However, a viable parameter space persists around $m_{\mathrm{DM}} \sim 60\,\mathrm{GeV}$ with couplings $\lambda \sim 10^{-3} - 10^{-5}$, where both models exhibit comparable behavior.
    \item For $m_{\mathrm{DM}} = 60\,\mathrm{GeV}$, the predicted neutrino fluxes in the NLHP and SM+Scalar models differ by no more than one order of magnitude, making discrimination challenging in this mass range.  
    \item At higher masses ($500\,\mathrm{GeV}$ and $1\,\mathrm{TeV}$), the NLHP model produces neutrino fluxes that can vary by up to five orders of magnitude depending on the non-linear couplings. This stark contrast highlights the NLHP's richer phenomenology and its potential to generate distinctive signatures in future neutrino telescopes.
    \item The non-linear structure of the NLHP, characterized by additional effective couplings, leads to significant deviations from the SM+Scalar predictions. These deviations are most pronounced at higher DM masses, suggesting that neutrino telescopes like IceCube-Gen2 and KM3NeT could play a pivotal role in probing the nature of the underlying symmetry-breaking mechanism.
    \item Our results underscore the importance of combining indirect detection with cosmological and direct detection constraints. While solar neutrino fluxes alone may not suffice to distinguish the models at low masses, their synergy with other probes could provide a more comprehensive test of DM scenarios.
\end{itemize}
In summary, solar neutrino observations offer a promising avenue to explore non-linear symmetry-breaking mechanisms in DM models. The NLHP model's enhanced sensitivity to high-mass DM candidates, coupled with its distinct neutrino flux predictions, positions it as a compelling framework for future indirect detection efforts. Further refinements in neutrino propagation simulations and detector sensitivity analyses will be essential to fully exploit this potential.

\acknowledgments

Authors L. L.-L. and A. C.-M. would like to thank GALC for its invaluable contributions to this work.

% Bibliography

%% [A] Recommended: using JHEP.bst file
 \bibliographystyle{JHEP}
 \bibliography{main}

\end{document}